\documentclass{aa}
\usepackage{graphicx}
\usepackage{color}

\def\aj{AJ}
\def\apj{ApJ}
\def\apss{Astroph.Sp.Sci.}
\def\aap{A\&A}
\def\mnras{MNRAS}

\begin{document}

\sloppypar

%

   \title{Broad band variability of SS433: Accretion disk at work?}

   \author{M.~Revnivtsev \inst{1,2}, S.~Fabrika \inst{3}, P.~Abolmasov 
\inst{4}, K.~Postnov\inst{4,5},  I.~Bikmaev \inst{6}, R.~Burenin \inst{2},
   M.~Pavlinsky \inst{2}, R.~Sunyaev \inst{1,2}, I.~Khamitov \inst{7}}

   \offprints{mikej@mpa-garching.mpg.de}

   \institute{
              Max-Planck-Institute f\"ur Astrophysik,
              Karl-Schwarzschild-Str. 1, D-85740 Garching bei M\"unchen,
              Germany,
      \and
              Space Research Institute, Russian Academy of Sciences,
              Profsoyuznaya 84/32, 117810 Moscow, Russia
         \and
              Special Astrophysical Observatory, Nizhnij Arkhyz,
Karachaevo-Cherkesiya, 369167,  Russia
         \and
              Sternberg Astronomical Institute, 119992, Moscow, Russia
          \and
        University of Oulu, Finland
         \and
              Kazan State University, Kremlevskaya
            str.18, 420008, Kazan, Russia
    \and
           TUBITAK National Observatory, Akdeniz Universitesi Yerleskesi, 07058, Antalya, Turkey
            }
  \date{}

        \authorrunning{Revnivtsev et al.}

   \abstract{We present broad band power spectra of variations of SS433
in radio, optical and X-ray spectral bands. We show that at frequencies
lower than $10^{-5}$ Hz the source demonstrates the same  variability pattern
in all these bands. The broad band power spectrum can be
fitted by one power law down to frequencies $\sim 10^{-7}$ Hz with
flattening afterwards. Such a flattening means that on time scales
longer than $\sim 10^{7}$ sec the source variability becomes
uncorrelated. This naturally leads to
the appearance of quasi-poissonian flares in the source
light curve, which have been regularly observed in radio and optical
spectral bands.
The radio flux power spectrum appears to have a
second break at Fourier frequencies $\sim 10^{-5}$ Hz which can be caused by
the smearing of the intrinsic radio variability on timescale of the 
light-crossing time of the radio emitting region. 
We find a correlation of the radio and optical fluxes of SS433 
and the radio flux is delayed by about  $\sim 2$ days with respect
to the optical one. 
Power spectra of optical and X-ray variabilities
continue with the same power law from $10^{-7}$ Hz up to $\sim 0.01-0.05$
Hz. The broad band power spectrum of SS433 can be interpreted in terms 
of self-similar accretion rate modulations in the accretion disk 
proposed by Lyubarskii (1997) and elaborated by Churazov et al. (2001). 
We discuss a viscous time-scale in the accretion disk of SS433 in
implication to the observed broad band power spectrum.  
   \keywords{accretion, accretion disks--
                black hole physics --
                instabilities --
                stars:binaries:general --
                X-rays: general  --
                X-rays: stars
               }
   }

   \maketitle

%

\section{Introduction}

SS433 is a high-luminosity massive X-ray binary system with
steadily precessing accretion disk and jets
(\cite{margon84}, see \cite{fabrika04} for a recent review).
Most of the system luminosity ($\sim 10^{40}$ erg/s) is emitted
in optical and UV spectral bands  (Cherepashchuk
et al., 1982; Dolan et al., 1997). The compact source in this binary 
system, probably a black hole, accretes the material from the 
companion late A-supergiant (Gies et al. 2002; Hillwig et al. 2004;
Cherepashchuk et al. 2003; 2004) at a highly super-Eddington rate.
Mildly relativistic ($v\sim0.26$c) baryonic jets are launched from the
innermost parts of the supercritical accretion disk.

The system demonstrates several periodicities (Cherepashchuk 2002): 
the precessional one ($\sim$162 days), the orbital one ($\sim$13 days), and
the nutational one ($\sim 6$ days). Studies of these periodicities
made it possible to tightly constrain some of the
system's parameters (Eikenberry et al. 2001; Collins \& Scher 2002)
including the binary system inclination angle ($i\sim78^\circ$) 
and the jet precession angle ($\sim20^\circ$).

X-ray emission produced in the innermost regions of the accretion flow is
completely screened from the line of sight by the optically and 
geometrically thick inner disk and outflowing wind.  
The observed X-ray flux from SS433 can be modeled by  
thermal bremsstrahlung emission from a $\sim 20$ keV plasma moving in the jets
(Watson et al. 1986; Kotani et al. 1996). A detailed analysis of
high resolution X-ray spectra taken by CHANDRA (Marshall et al. 2002)
basically confirms previous findings.
The recent analysis of hard X-ray spectra of SS433 obtained by
INTEGRAL, however, indicates that a mixture of plasmas with 
different temperatures and geometries could be present in the 
jet region (Cherepashchuk et al. 2003; 2004).

The source demonstrates variability in all spectral bands and on various
time scales (see e.g. \cite{fiedler87}, Goranskii et al. 1987,
Zwitter et al. 1991, Kotani et al. 2002). However, thus far
no systematic studies of aperiodic variability of SS433
at different wavelengths have been carried out.
Previous works were mainly focused on the analysis of periodic 
variabilities found in the source.
SS433 has been regularly monitored at different wavelengths
over the last decades and appears to be also well suitable for analysis
of aperiodic variations. The long-term optical observations indicate that
the source demonstrates on average a very stable behavior with
no drastic changes in its activity pattern (Eikenberry et al. 2001;
Goranskii et al. 1998).
It is the only known persistent galactic microquasar
with nearly constant kinetic power in jets (Fabrika 2004).

An enhanced radio flux is observed from the source at some periods.
These 
were classified as active states on top of the quiescent (passive) states
(Bonsignori-Facondi et al. 1986; Fiedler et al. 1987; Trushkin et al. 2003).
However, on the long range, it can be shown that the appearance of the
active radio states in SS433 is aperiodic process and can be well described
in the language of power spectra.

In this paper we for the first time attempt to systematically study the
aperiodic variability of SS433 by constructing power spectra of its flux
variations in radio, optical and X-ray spectral bands. We found that
variabilities at different wavelengths can be described by one power law
spanning several orders of magnitude in frequency, which is suggestive of
their common nature. We discuss the possible origin of the obtained power
spectra as being due to the mass accretion rate modulation in the underlying
accretion disk.

\section{Observations and data analysis}

To estimate power spectra of the source variability
in different spectral bands we have combined all data available to us.
As the data were obtained by different instruments, are not evenly spaced
 and have significant time gaps,
we can not use the simple discrete Fourier transform, which is usually applicable
for evenly spaced 
data (e.g \cite{leahy83}).

The optical and radio data are analyzed by means of
periodograms that provide
estimations of the Fourier amplitude at a given Fourier
frequency $f$ in the form
(\cite{deeming75,lomb76,scargle82}):

\[
|a(f)|^2=\left[\sum_{i=1}^{N}f(t_i)\cos(2\pi f t_i)\right]^2+\left[\sum_{i=1}^{N}f(t_i)\sin(2\pi f t_i)\right]^2
\]
Here $f(t_i)$ is the flux measurement at a given time $t_i$ with subtracted
average flux value, $N$ is the number of measurements in the set.
To obtain the fractional rms squared
normalization we calculate the power $P(f)$ in the form:

\[
P(f)={2T\over{\langle f\rangle ^2 N^2}}|a(f)|^2
\]
where $T$ is the time span of the set ($T=t_N-t_1$).
The power spectrum calculated with this
normalization has one important property:
The integral of the power spectrum from $f_1$ to $f_2$ gives the square of
fractional rms variability which is present in the lightcurve
of the source on time scales from $f_2^{-1}$ to $f_1^{-1}$.

The longest time series were divided into a few parts and the power
spectrum was evaluated for each of them. Estimates of the power of different
parts lying at close Fourier frequencies were averaged and the
standard deviations of these values were calculated.

For all our data sets we have made simulations in order to check the effect of
uneven sampling. Simulations show that this effect distorts
the power spectrum in the form
of a power law ($P\propto f^{-\alpha}$) at the low
frequency end, but not very significantly. In most cases such a distortion
does not exceed the uncertainty in the power estimates.

\subsection{SAI database of observations of SS433}

The optical broad band power spectrum of SS433 was calculated
using a data-base of optical V-band photometry of SS433 obtained in
1979--1996. The data-base consists of 2200 individual nights of
observations collected at Sternberg Astronomical Institute
(\cite{goranskii98}). In this data--base all published V-band
observations are reduced to one photometrical system. The average
uncertainty of the photometrical data in individual nights is
5-10\%. The data--base also includes series of photometric
observations taken during several consecutive nights (for example,
Goranskii et al. 1987; 1997), with each night containing from several
to a few tens individual observations. The contribution of
measurements uncertainties to the obtained power spectrum (the white
noise component) have been subtracted. Besides, we have removed
from the power spectrum the peaks due to coherent modulations with
orbital, nutational and precessional periods.

\subsection{RTT150}
The high-frequency power spectrum of optical fluctuation was calculated using
observations of SS433 performed with 1.5-m Russian-Turkish
Telescope (RTT150) at T\"{U}BITAK National Observatory (TUG), Bakyrly
mountain.
The observations were carried out in September 2004 during seven nights
under clear sky but poor seeing ($\approx 2''$) conditions. In our analysis,
we used points out of the orbital eclipse.
A low readout noise back-illuminated 2$\times$2\,K Andor Technologies
DW436 CCD mounted in F/7.7 Cassegrain focus of the telescope was used.
The photometric R filter was utilized because of lower Galactic absorption in
this filter and larger magnitude of SS433
(R$\approx 12$) than in blue bands. The method of fast photometry we used
is described in some detail in Revnivtsev et al. (2004)

\subsection{X-ray data: RXTE/ASM and EXOSAT/ME}

In order to construct the broad band power spectrum of variability of SS433
in X-ray band we have used two sets of data. Until now, the longest
time-series of SS433 in X-rays is provided by the All Sky Monitor
(http://xte.mit.edu/ASM\_lc.html) onboard the Rossi X-ray Timing Explorer
(\cite{rxte}).

The power spectrum calculated from the observed X-ray ASM light curve should
have the statistical noise contribution (a constant component in the power
spectrum of the source). The value of this constant component in the power
spectrum in principle can be calculated theoretically by using the values
of statistical uncertainties quoted in the light curve. However, the errors
quoted in the ASM light curves often underestimate the real uncertainties
of the measured fluxes (see e.g. \cite{grimm02}). Therefore in order to
estimate the contribution of the noise component we have studied the
power spectra of persistent weak sources, the supernova remnants
Cas A and Puppis A. The obtained noise component
after the appropriate renormalization was then subtracted from the power
spectrum of SS433.

To construct the X-ray power spectrum at frequencies $>2\times 10^{-5}$ Hz
we have used data of EXOSAT/ME observations. The EXOSAT observatory
had orbit much longer than that of RXTE thereby providing
much longer uninterrupted sets of observations. Preprocessed lightcurves
of SS433 were taken from the HEASARC archive at GSFC
(http://legacy.gsfc.nasa.gov). Only observations performed at precessional
phases $0.8<\psi<1.2$ were considered. At these precessional phases the
X-ray flux from jets of SS433 is maximal.
Power spectra were constructed using the
standard task of FTOOLS 5.2 package.

\subsection{Radio observations}

The longest available set of radio flux from SS433 at frequencies
2.25 GHz and 8.3 GHz is provided by the GBI monitoring
(http://www.gb.nrao.edu/ fgdocs/gbi/gbint.html). We have used
measurements at the frequency 2.25 GHz because these have higher
signal to noise ratio than those at 8.3 GHz. The contribution of
statistical noise was subtracted from the obtained power spectrum.
The available data allowed us to estimate the power at frequencies
from $\sim 5\times 10^{-9}$ Hz to $\sim 2\times 10^{-6}$ Hz and at
$\sim 10^{-4}$ Hz. Before the binning of the power spectrum we
have removed the peak of coherent variations with the nutational
period ($\sim$6 days, \cite{trushkin01}).

In order to extend the power spectrum of radio variability of
SS433 to higher frequencies we have included the estimate of the
SS433 variability on time scales of tens and hundreds of seconds
made by Rand et al. (1988). Note that these measurements were done
at the frequency 0.4GHz (different from 2.25 GHz for the GBI
measurements), so the relative normalization of the power
estimates might be different. However, comparison of available
long time scale measurements of radio flux of SS433 at 0.4 GHz 
(e.g. \cite{bf86}) with that at 2.25 GHz 
tells us that the normalization correction factor
between variabilities at these two frequencies does not exceed $\sim2-3$.

\begin{figure}[htb]

\includegraphics[width=\columnwidth,bb=33 166 570 715,clip]{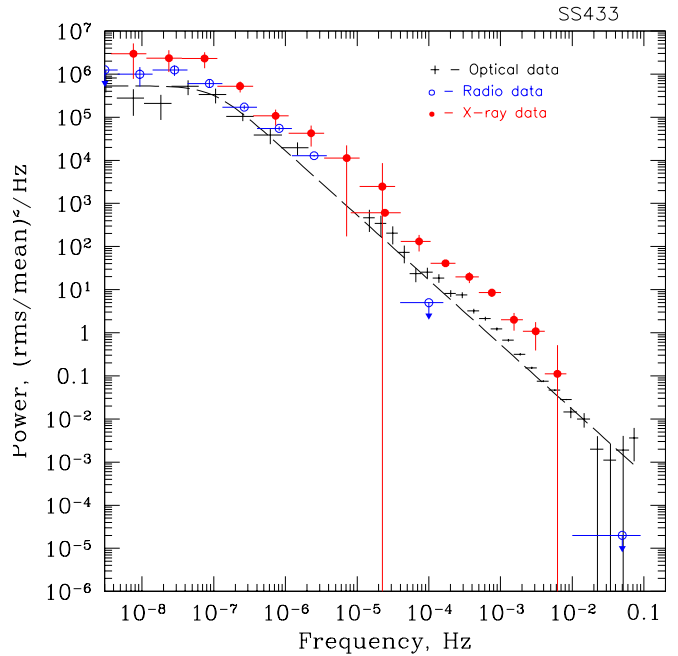}
\caption{The power spectrum of the SS433 variability in
different spectral bands. The optical power spectrum is shown by crosses, the radio power spectrum
is shown by open circles,
the X-ray power spectrum is represented by filled circles.
No special rescaling is done. The dashed
line shows the fit in the form $P(f)\propto (1.0+(f/f_{\rm break})^{\alpha})^{-1}$, where $\alpha=1.5$}
\label{power_total}
\end{figure}

\section{Results}

The resulting broad band power spectra of aperiodic variability of SS433 in
different spectral bands is shown
in Fig.\,\ref{power_total}. Two most important features of the
power spectra are clearly seen.
\begin{itemize}
\item
First: stochastic flux variabilities in all three energy ranges  
(radio, optical and X-ray) follow very similar patterns. The power spectra
in these ranges are flattened in Fourier frequencies $\la 10^{-7}$~Hz.
\item
Second: the power-law part of the optical and X-ray power spectra
(which can be traced up to
Fourier frequencies of the order of 0.01-0.1~Hz) spans over more
than five orders of magnitude in frequency.
The power spectrum in radio is likely to have a break between $10^{-6}$~Hz
and $10^{-4}$~Hz.
\end{itemize}

\subsection{Self-similar power spectrum -- is it produced by accretion disk?}

The self-similar behavior of variability of sources is observed not
for the first time. Decades ago it was noted that at low Fourier frequencies
($<0.01-0.1$Hz)
X-ray binaries demonstrate a self-similar power law component
in their power spectra (e.g.
\cite{vdk87,hvdk89}). A detailed study of variability of the black hole binary
system Cyg X-1 in the soft spectral state (when the optically thick accretion
disk presumably extends down to the last stable orbit around the black hole)
have demonstrated that this power-law component can extend
over four orders of magnitude
in Fourier frequency (\cite{chur01}). An extensive study of low mass X-ray
binary systems revealed that such power-law behavior is observed in practically
all LMXBs (Gilfanov \& Arefiev 2005).

It was suggested that the observed self-similar variability in X-ray
binary systems is a result of mass accretion rate variations in the accretion
disk around the compact object 
(\cite{lyubarski97, chur01}, Gilfanov \& Arefiev 2005).
The observed variability in X-ray binaries
at all Fourier frequencies is rather strong (a significant part of the total
luminosity varies), which implies that inner
regions of the accretion disk, where the most of the energy release takes
place,
are involved. These regions, in turn, respond to the accretion rate
modulations which have diffused there from the outer parts of the disk.
If modulations of the accretion rate occur at any given radius of the
accretion disk, imprinting the characteristic frequency
to the inflowing matter, one can readily get the self similar power
spectrum of the form $P\propto f^{-\alpha}$, where $\alpha\sim 1-2$
for a wide range of accretion disk models (\cite{lyubarski97}).

\subsection{Viscous time scale in the accretion disk}

Such a power-law shape of the accretion disk power spectrum is expected to extend up to time scales on which the disk variability becomes
 uncorrelated, i.e.
up to the longest time scale of the accretion disk
(\cite{chur01}, Gilfanov \& Arefiev 2005).
From theoretical point of view this time scale is the time of viscous
diffusion of matter from the outermost regions of the accretion
disk to the innermost parts. For example,
in the framework of standard optically thick accretion disk 
(\cite{ss73}, see also \cite{lbp74}):

\[
t_{\rm visc}\sim \left({H\over{R}}\right)^{-2} {T_K(R)\over{\alpha}}
\]
where $H$  is the hydrostatic height of the disk at radius $R$,
$T_K(R)$ is the Keplerian rotation period at this radius, $\alpha$
is the viscosity parameter.

At frequencies below $\sim t_{\rm visc}^{-1}$ the variability
becomes uncorrelated and consequently the power spectrum should
flatten. As mentioned above, this indeed was found to be the case for
all LMXBs (Gilfanov \& Arefiev 2005). 
The uncorrelated 
variability at frequencies $f< t_{\rm visc}^{-1}$  
should appear on the light curves
of the source in different energy bands as a set of random flares
with durations  around $\sim t_{\rm visc}$.
 Such ``flares'' have been actually often observed
in radio and optical spectral bands,
which led to the classification of the
so-called ``active'' and  ``passive'' states of SS433 (e.g. \cite{bf86}, 
\cite{irsmambetova97}, \cite{fabrika04}).
Such a flaring behavior in reality appears to be
 a natural observational appearance of the
 uncorrelated variability 
on time scales larger than the longest (viscous) time 
 scale of the accretion disk.
Note that a very similar phenomenon (but
on much shorter time scales)
has been observed many times from the best studied black hole
candidate Cyg X-1,whose  variability can be described as a series 
of random shots
(the ``shot noise'' model, e.g. \cite{terrell72}, \cite{vihl94}, 
\cite{negoro94})

The break in the power spectrum of SS433 at $f \sim 1-2\times 10^{-7}$ Hz
corresponds to a time scale of 50-100 days. The ratio of
this time scale to the orbital period 13.6 days in SS433 is $\sim 5-
10$, which is in a reasonable agreement with results obtained by
Gilfanov \& Arefiev (2005) for LMXBs.
\footnote{Note here that the ratio of orbital to the break frequency of the 
order of 5-10 may indicate that the mass ratio in the binary system SS433 is 
larger than $q>0.3$ (see results of Gilfanov \& Arefiev 2005)}
 Such a low value of the
ratio may indicate that the outer accretion disk is relatively thick
$H/R\ga 0.1-0.2$ (see also the discussion in Gilfanov \& Arefiev
2005). In the case of SS433 the thickness of the outer parts of
the disk should be at least not smaller than that found 
in LMXB systems because
of the very high accretion rate. So in principle for a low alpha-parameter
$\sim 0.1$ the viscous time
scale of the accretion disk in SS433 evaluated from the above formula 
can be comparable to the value derived from the power spectrum analysis.

Note here that the ratio of orbital to the break frequency of the order of
5-10 may indicate that the mass ratio in the binary system SS433 is larger 
than $\sim 0.35$ (see results of Gilfanov \& Arefiev 2005).

However, there are observational evidences of the large 
thickness of the disk in SS433. The approximate equality of the amplitudes of
the precessional modulation and of the primary eclipses (the latter 
measured in the precessional phases when the disk is maximum open) in
optical bands (\cite{fabrika02, fabrika04}) indicates that the 
size of the donor is almost equal to that of the geometrically 
thick disk. The same relationship was found in different X-rays bands 
(Cherepashchuk et al. 2004): from 1.5~keV to 100~keV both the 
precessional and eclipsing amplitudes increase while keeping comparable 
to each other.
This means that for all acceptable binary mass ratios in SS433 
($q = M_{x}/M_{op}=0.2 - 0.5$) both the donor and 
the outer disk rim sizes are about $\sim 0.4 - 0.5$ in units of the binary 
separation and the estimate $H/R\sim 1$ looks quite plausible. 
This may be a point of concern. 
For a disk with $H/R\sim 1$ the long viscous time $\sim 10 T_K$ 
(recall that $T_K$ relates to the Keplerian
time at the outer disk radius) seems to be unlikely unless a
small effective viscosity is assumed. 

So to have a long viscous time scale in the accretion disk in SS433 
we can admit that either some outer
parts of the disk (through which matter diffuses on 
the longest time scales) have a moderate thickness 
$H/R\sim0.2-0.3$, or the viscosity parameter $\alpha$ in the
accretion disk of SS433 is smaller than 0.1

The viscous time scale in an accretion disk can be treated as the
time it takes for the accreting material to pass through the disk.
This time may be estimated
from the analysis of the nutational variability of jets in SS433
(see e.g. \cite{katz82,collins86,fabrika04}). The idea is that 
the nodding motions of the jets and the inner accretion disk 
have to be delayed with respect to 
tidal perturbations of the outer parts of the disk 
induced by the secondary  
companion. This 
time-delay was estimated to be of the order of a day. This means
that the time for the passage of material through the disk is either
extremely short (practically the free-fall time) or it is an integer 
multiple of the 
nutational period because of the 
periodic nature of the nodding motions.   

A short travel time of the material passing through the disk is  
in strong disagreement
with our estimates inferred from the noise power spectrum. 
We should note here
that the estimate of the travel time from the nutational variability
actually measures the propagation time of tidal perturbations from
the outer parts of the accretion disk to the inner parts. Periodic
tidal forces from the optical component may form a standing (in
the binary's reference frame) structures in the outer accretion
disk (see e.g. model calculations  in \cite{blondin00} and
references to earlier works therein) which can translate the
nutational perturbations from outer parts of the disk to its inner
parts on time scales much shorter than the matter travel time
through the disk. In this picture, 
tidal perturbations may propagate through the disk with velocities 
close to the sound velocity.

 Of course, the disk in SS433 is far from being the standard one. 
It is strongly supercritical and is inclined to the orbital plane
($\approx 20^{\circ}$).
However, we can use as a guide the analogy with the twisted tilted
subcritical accretion disk observed in accreting binary system 
Her X-1. In that system the observed X-ray flux is strongly modulated
with a period of $\sim 35$ days. 
The binary inclination in Her X-1 is close
to 90 degrees, so we observe the system nearly edge-on. 
Over the 35-day period the X-ray emission appears twice -- the bright 
``main-on'' state and then a weaker ``low-on'' state.
In Her X-1 the observed 35-day modulation
can be understood as the precessional motion of the outer parts 
of the inclined accretion disk (Gerend \& Boynton 1976). 
The analysis of the beginning of an X-ray cycle (the so-called turn-on
phase) observed by RXTE (Kuster et al. 2005) confirms that it is the outer
edge of
the accretion disk that opens the central X-ray source at the turn-on.
The disk is not flat, it is warped (or
twisted) 
because of the strong interaction of the innermost parts with the rotating
magnetosphere of the central neutron star in Her X-1. The inner parts of
the disk start screening the central X-ray source at the end of the main-on
X-ray state (see e.g. Shakura et al. 1999 and references therein).
The disk in Her X-1 precesses as a whole
in the direction opposite to the orbital revolution, so the inner
parts of the disk appear to lead the outer parts in the precessional motion.
This can be realized only when the tidal interaction from the 
outer parts is translated
to the inner parts of the disk on a time scale which is shorter
than the viscous time of the disk (of order of 20-30 days).

\subsection{Accretion disk imprints in SS433 variability}

{\it Optical}. It can be considered well established observationally
that most of the optical variability in SS433 comes from the 
so-called accretion disk
funnel, the open cone in the innermost part of a supercritical
accretion disk (see e.g. \cite{fabrika04} for a review and 
recent calculations by Okuda et al. 2004). 
The funnel itself may by partially hidden from the direct view by
the strong wind from inner accretion disk. 
In this case some fraction of the optical radiation 
and variability will come from the wind photosphere. Both the funnel 
and the inner wind surround the bases of the jets. The structure of this
region is not a subject of this paper, the most important thing here is that
the energy emitted in optical and UV bands is very large, of the order
of $\sim 10^{40}$ erg/s. This undoubtedly tells us that most of the
optical emission should be originating from reprocession of the
internal X-rays generated in the inner accretion disk. 
Therefore in the framework
of self-similar variability of the mass accretion rate in the disk
(\cite{lyubarski97, chur01}) the optical variability of SS433 should
share the same time properties as observed in X-ray range. It 
naturally should have the self-similar power spectrum from $f\sim
t_{\rm visc}^{-1}$  to $t_{\rm lc}^{-1}$, where $t_{\rm visc}$ is
the viscous time at the outer boundary of the accretion disk (the longest
time scale in the disk) and $t_{\rm lc}$ is
the light crossing time of the funnel. At frequencies above
$f\sim t_{\rm lc}^{-1}$ the optical photometric variability should
be smeared out.

In SS433 the size (height) of the funnel can be evaluated to
be of the order of $10^{12}$ cm 
(\cite{heuvel81}, \cite{marshall02}, \cite{mikej04}) and therefore
the limiting frequency $f\sim t_{\rm lc}^{-1}\sim 0.01$ Hz.
Unfortunately, the existing data do not allow us
to detect the decline of the power spectrum
at frequencies higher than $\sim 0.01$~Hz (see Fig. \ref{power_total})
but future observations with higher statistics 
may reveal it.

It is interesting to note here that the variability of optical 
emission lines, which are emitted in the jets at distances above  
$\sim 10^{14}$~cm (Borisov \& Fabrika 1987) from
the compact object where the jet cross size is of the same order,
should start washing out at the jet light-crossing frequencies of the order
of $\sim 10^{-4}$ Hz, which should be detectable.

{\it X-ray.}
The X-ray flux of SS433 originates at the base of hot outflowing jet
formed in the innermost regions of the accretion flow.
Therefore we can naturally anticipate the X-ray flux to contain
information about the
mass accretion rate variations present in the innermost parts of the
disk which are presumably brought in from the outer parts of the disk
with the accreting matter. In this picture the similarity of the
power spectra of variability of SS433 in optical and X-ray spectral band
is quite natural. Indeed, the direct correlation of variability of SS433
in optical and X-ray bands was recently discovered by Revnivtsev et al. (2004).

Note that the X-ray variability should also be smeared out at high frequencies
(in X-rays we observe the cooling jets). The limiting frequencies
depend directly on the X-ray jets length $r_j$, which presumably 
is $\sim 10^{12}$~cm (\cite{w86}, Kotani et al. 1996). So a decline
in the X-ray power spectrum at frequencies higher than $0.01 - 0.1$~Hz is
expected.

\begin{figure}[htb]

\includegraphics[width=\columnwidth,bb=33 186 570 520,clip]{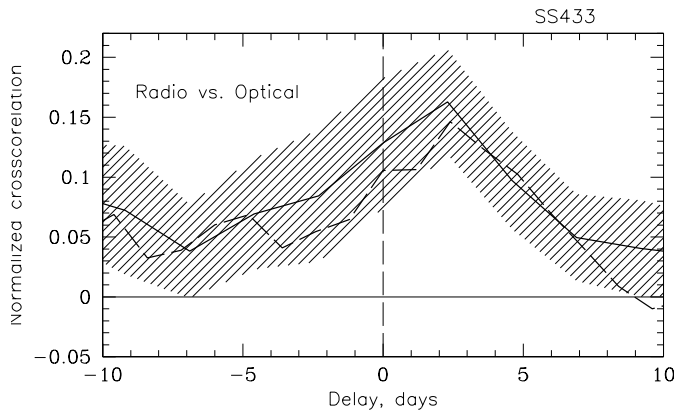}
\caption{The cross-correlation of radio ($\sim$2 GHz) and optical 
(photometric) variability. The light curves were rebinned into 
$2\times10^{5}$ sec time bins. The positive lag means that the radio 
flux is delayed with respect to the optical one.
The shaded region denotes the rms uncertainties in the measurements
of the cross-correlation at any given time lag (estimated from the data). 
The solid line shows the average cross-correlation function. 
The dashed line represents the average cross-correlation measured 
from the light curves rebinned into $10^{5}$ sec bins}
\label{crosscor}
\end{figure}

{\it Radio.}
Most radio emission observed from SS433 originates in the jets.
The power spectrum obtained by us at very low Fourier frequencies 
($<10^{-5}$~Hz,
see Fig. \ref{power_total}) clearly indicates that variability of radio
emission in SS433 has the same origin as the optical and X-ray variability.
Even more direct confirmation of this hypothesis can be extracted
by cross-correlating the optical and radio time series.
From Fig. \ref{crosscor} we can see that the light curves at these
spectral bands are significantly correlated.
The radio flux is delayed with respect to the optical
photometric variability with a time lag of $\sim$2 days. This
time lag approximately corresponds to the jet travel time
from the accretion disk funnel (where most photometric variability is
generated, see above) to the distance where the radio emission appears, 
$\sim 10^{15}$~cm (Paragi et al. 1999).

The observed power spectrum (Fig.~1) indicates that there is no
additional generation of the radio flux variability in jets on longer time scales
($>10^{5}$ sec). The points on the power spectrum at frequencies
above $\sim10^{-5}$ Hz show that radio flux have a smaller
variability amplitude than that found in the optical and X-rays
at same frequencies.
This can be a result of strong smearing of any possible
intrinsic variability in the radio flux on time scales shorter
than the light-crossing time of the radio emitting region.

\section{Discussion}

As we have shown above, the straightforward interpretation of
the obtained broad-band power spectra of aperiodic flux variations
in SS433 can be made in terms of self-similar accretion rate
modulations by the disk. How solid is this interpretation? It
is based on the assumption that the observed frequency of
the flattening of the power spectrum
can be associated with the viscous time scale of the accretion disk.
This assumption seems to be supported by the analysis of power spectra
of a dozen of LMXBs made by Gilfanov and Arefiev (2005).

The accretion disk in SS433 is a complex structure and clearly is
different from disks in LMXBs. A time-dependent hydrodynamic model 
of an accretion disk in a close binary (Blondin 2000) shows that the tidal
forces truncate the accretion disk to radii of order half the average 
radius of the Roche lobe and an effective $\alpha$--parameter is of order
0.1 near the outer edge of the disk. The disk in SS433 has to be even  
more complex, it is inclined by an angle of 20 degrees to the orbital plane. 
So the resulting thickness of the
disk must be further increased in comparison with 
any estimates based on the standard accretion disk theory.

From the observational point of view there are clear indications
that the inner disk around compact object in SS433 should be 
very thick ($H/R\sim$1). If the $\alpha$ parameter of viscosity 
in such a disk is $\alpha\sim0.1$ the disk viscous
time scale can not be as large as $\sim$100 days.
To obtain such a long viscous time, 
which are necessary to ensure 
the long time scale modulations of the mass accretion
rate, 
we can assume that there is some (probably small) 
part of the outer disk with a moderate disk thickness $H/R\sim0.1-0.2$

A possible alternative explanation can be delineated as follows.
Let the viscous time of the disk be actually small, of order of
$T_K \sim$ a few days. The short viscous time scale is
favored from the nutational periodicity analysis. The disk will
generate the power-law modulations in the accretion rate all the
way down to the frequency $1/T_K\sim 10^{-5}-10^{-6}$ Hz
irrespective of what kind of fluctuations are originally supplied
with the incoming matter from the companion. So to explain the
same power-law continuation of the spectrum to lower frequencies,
as observed, we need to invoke some physical mechanism of
generation of the power-law fluctuation spectrum from the donor star.
In principle, such a mechanism could be a feed-back between the
optical/UV flux fluctuations in the luminous 
central source (and the disk wind) and the
illuminated photosphere of the donor, which will be
transferred to the outer disk region with matter passing through
the vicinity of the inner Lagrangian point in the binary system. 
Indications of a strong heating effect in the donor of SS433 have 
been recently found through optical spectroscopy 
of the companion A-supergiant carried out at the precessional phases close
to the maximum disk opening
(Cherepashchuk et al. 2004). 
Depending on orbital and precessional orientations (phases) 
different parts of the donor's surface should be exposed for the heating. 

The self-similar behavior of fluctuations is quite common in various
natural phenomena, and the power-law noise spectrum 
can be generated in the
turbulent disk outflow (e.g. Zeleny \& Milovanov 2004 for a
recent review). In this picture,
the lower limit frequency of correlated fluctuations can be associated
with star's precession frequency $f_{pr}\sim 1/160d\sim 6\times
10^{-8}$ Hz, which does not contradict to the observed flattening
frequency. However, the inspection of the obtained power spectrum
does not reveal a significant break at frequencies about $10^{-5}$~Hz 
corresponding to the short viscous time scale of the disk.

So the simplest approximation of the obtained fluctuation power 
spectrum as one power law from $10^{-7}$ to $10^{-2}$ Hz appears to 
support the long viscous time scale of the accretion disk 
in SS433 and the interpretation 
of this spectrum in terms of self-similar modulations of the 
accretion rate in the disk.

\section{Summary}

We constructed the broad band power spectra of aperiodic variability of
high-mass X-ray binary system SS433 in radio, optical and X-ray spectral
bands. The obtained properties of the spectra 
can be summarized as follows:

\begin{itemize}
\item
Power spectra of variability of SS433 in radio, optical and X-ray spectral
bands have the power law shape ($P\propto f^{-\alpha}$) from 
$\sim 0.1$ Hz down to $10^{-7}$~Hz
with flattening afterwards. At frequencies $>10^{-7}$~Hz the power law index
is approximately $\alpha \sim 1.5$

\item The power spectrum of radio variability presumably has one more
break (steepening) at frequencies $\sim 10^{-5}$~Hz which can be a result of
smearing with large light-crossing time of the radio emitting region
($\sim10^{5}$ sec)

\item Radio and optical (photometric) fluxes of SS433 are correlated
and the radio flux is delayed by about  $\sim 2$ days with respect
to the optical one.
The correlated variability in optical and X-ray spectral bands was
previously detected by Revnivtsev et al. (2004)
\end{itemize}

The observed behavior of fluxes of SS433 in different spectral bands can be
interpreted in the framework of self-similar accretion disk variations
proposed by Lyubarskii (1997) and elaborated by Churazov et al. (2001)
and Gilfanov \& Arefiev (2005).
The accretion disk introduces the mass accretion rate variations at any given
radius for example due to fluctuations of the viscosity parameter
$\alpha$. The resulted variable mass accretion rate
enters the region of the maximum energy release near the compact object
and generates powerful variations in the emitted X-rays which then
are reradiated in optical and UV spectral bands by a funnel 
surrounding the jets.
We do not see directly this X-ray emission from the innermost parts 
of the accretion flow because of the high inclination of the binary 
system and a large optical depth of the geometrically thick accretion 
disk and the outflowing wind. Mass accretion rate fluctuations 
are translated to the density variations in the outflowing hot 
baryonic jet and give rise to the observed X-ray variability
correlating with the optical one (Revnivtsev et al. 2004).
As the expanding jet cools down, it starts
emitting in optical lines and 
in radio. The time lag between the
radio flux and the optical photometric flux (emission of the accretion disk
funnel) equals to the jet travel time from the funnel to the radio
emission region.

\begin{acknowledgements}
The authors acknowledge N.I.Shakura for discussions.  
This work was partially supported by grants of
Minpromnauka NSH-2083.2003.2, NSH-1789.2003.02 and program of Russian
Academy of Sciences ``Non-stationary phenomena in astronomy''.
Partial support through RFBR grants 03-02-16110, 04-02-16349 and 02-02-17174
is acknowledged. PK acknowledges the University of Oulu for hospitality and
support through grant of Academy of Finland 100488.
MR,IB,RB and MP thank International Space Science Institute (ISSI, Bern, Swiss)
for partial support.
Research has made use of data obtained from
High Energy Astrophysics Science Archive Research Center
Online Service, provided by the NASA/Goddard Space Flight Center.
Also data from GBI-NASA monitoring program were used. The Green Bank
Interferometer is a facility of the National Science Foundation operated
by the NRAO in support of NASA High Energy Astrophysics programs.

\end{acknowledgements}

\end{document}